\documentclass[prl,onecolumn]{revtex4}
\usepackage{graphicx}

\begin{document}

\title{Experimental Preparation of Quadripartite Cluster and GHZ Entangled States
for Continuous Variables}
\author{Xiaolong Su, Aihong Tan, Xiaojun Jia*, Jing Zhang, Changde Xie, Kunchi Peng}
\address{The State Key Laboratory of Quantum Optics and Quantum Optics Devices,\\
Institute of Opto-Electronics, Shanxi University, Taiyuan, 030006,
P.R.China}

\begin{abstract}
The cluster states and Greenberger-Horne-Zeilinger (GHZ) states are
two different types of multipartite quantum entangled states. We
present the first experimental results generating continuous
variable quadripartite cluster and GHZ entangled states of
electromagnetic fields. Utilizing four two-mode squeezed states of
light and linearly optical transformations, the two types of
entangled states for amplitude and phase quadratures of light are
experimentally produced. The combinations of the measured quadrature
variances prove the full inseparability of the generated four
subsystems. The presented experimental schemes show that the
multipartite entanglement of continuous variables can be
deterministically generated with the relatively simple
implementation.
\end{abstract}
 \maketitle
In recent years the investigation on continuous variable (CV)\
quantum communication network (QCN) has attracted extensive interest
along with the development of the experimental performances of
quantum information based on CV tripartite
Greenberger-Horne-Zeilinger (GHZ) entanglement\cite {one,two,three}.
A variety of complicated CV QCNs exploiting multipartite GHZ
entangled states of more than three subsystems have been
theoretically proposed\cite{four,five,six,seven}. Besides, an other
type of CV multipartite entangled states, CV cluster states have
been theoretically introduced\cite{eight}. Successively, Menicucci
et al. prove that the universal quantum computation (QC) can be
achieved with CV cluster states as long as a non-Gaussian
measurement can be performed\cite{nine}. In the study
of quantum information for discrete variables, the five-photon GHZ states%
\cite{ten} and four-photon cluster states\cite{eleven} have been
experimentally demonstrated and successfully applied in
open-destination teleportation and one-way quantum computing,
respectively. However, the realization of CV cluster and
quadripartite GHZ states still remain an experimental challenge. The
CV quantum resources, such as squeezed and entangled states of
light, emerge from the nonlinear optical interaction of a laser with
a crystal in an unconditional fashion. The unconditionalness of the
CV implementations makes the CV approach to be particularly suited
for further experimental demonstration of the general principles of
GHZ-state QCN and cluster-state QC.

In this letter, we present the experimental schemes to produce CV
quadripartite entangled cluster-state and GHZ-state. Using a pair of
non-degenerate optical parametric amplifier (NOPA) we obtained two
amplitude--quadrature squeezed states and two phase-quadrature
squeezed states, simultaneously. Then, only by means of linearly
optical transformation of these squeezed lights under certain phase
relations the CV quadripartite cluster-state and GHZ-state of
electromagnetic field are generated from the system, respectively.
The results measured with the balanced homodyne detectors to the
variances of amplitude and phase quadratures demonstrated that the
combinations containing both conjugate variables of all four modes
satisfy the criteria of the fully inseparability of
multimodes\cite{twelve}.

It has been theoretically and experimentally demonstrated that the
two coupled modes of the original signal and idler modes with
orthogonal polarizations from a NOPA are the phase-quadrature and
the amplitude-quadrature squeezed states of light, respectively\cite
{thirteen,fourteen,fifteen}. For a NOPA operating at deamplification
(the pump light and the enjected signal light are out of phase), the
superposed mode at $+45^0$ polarizing direction is the quadrature
amplitude-squeezed state and that at $-45^0$ is the quadrature
phase-squeezed state\cite {thirteen,sixteen}. The schematic of the
experimental system is shown in Fig.1. The pump laser (Nd:YAP/KTP)
is a homemade cw intracavity frequency-doubled and
frequency-stabilized Nd-doped YAlO$_3$ perovskite laser with a
frequency-doubling crystal KTP (potassium titanyl phosphate) inside
the cavity. The second harmonic wave output at $540nm$ and the
fundamental wave output at $1080nm$ from the laser are used for the
pump fields and the injected signals of the two NOPAs (NOPA1 and
NOPA2), respectively. The two NOPAs were constructed in identical
configuration, both of which consist of an $\alpha $-cut type-II KTP
crystal and a concave mirror. The front face of the KTP was coated
to be used as the input coupler and the concave mirror as the output
coupler of the squeezed states, which was mounted on a
piezo-electric transducer (PZT) for locking actively the cavity
length of NOPA on resonance with the injected signal at 1080nm.
Through a parametric down conversion process of type-II phase match,
the two-mode squeezed states of light at 1080nm were produced. The
output optical field from NOPA1 (NOPA2) is splitted by the
polarizing-beam-splitter $PBS_1$ ($PBS_2$) to the amplitude-squeezed
state $a_2$ ($a_3$) and the phase-squeezed state $a_1$ ($a_4$). The
half-wave plate $P_1$ and $P_2$ are
used for orienting the polarizations of the light beams toward $PBS_1$ and $%
PBS_2$, respectively. The quadrature amplitudes ($X_{ai}$) and phase ($%
Y_{ai} $) of the four squeezed modes $a_i$ ($i=1,2,3,4$) are expressed by%
\cite{fifteen,seventeen}

\begin{eqnarray}
X_{a1(4)} &=&e^{+r}X_{a1(4)}^{(0)},\text{ \quad }%
Y_{a1(4)}=e^{-r}Y_{a1(4)}^{(0)}, \\
X_{a2(3)} &=&e^{-r}X_{a2(3)}^{(0)},\text{ \quad }%
Y_{a2(3)}=e^{+r}Y_{a2(3)}^{(0)}.  \nonumber
\end{eqnarray}
Here, $r$ is the squeezing parameter, which depends on the strength
and the time of parametric interaction in NOPA, and we have assumed
that $r$ for the four squeezed state is identical for simplicity and
the requirement is easy to be reached in the experiments if the two
NOPAs were constructed in identical configuration and the
intracavity losses of the four modes were balanced. The values of
$r$ can be from zero to infinite ($r=0$ no squeezing, $r\rightarrow
+\infty $ ideal squeezing and the ideal limit can
not be achieved experimentally since it requires infinite energy). $%
X_{ai}^{(0)}$ and $Y_{ai}^{(0)}$ stand for the quadrature amplitudes
and phases of the injected signal fields into NOPAs. In experiments
we made all injected quadratures equal and normalized them to the
shot noise limit (SNL) of total four modes (taking the fluctuation
variance $<\Delta ^2(X_{ai}^{(0)})>=<\Delta ^2(Y_{ai}^{(0)})>=\frac
14$) in the following
calculations. $BS_i$ ($i=1,2,3$) is the $50\%$ beam-splitter and $PZT_i$ ($%
i=1,2,3$) is the piezo-transducer. At first interfering modes $a_2$
and $a_3$
on $BS_1$ with the phase difference of $\pi /2$ which was controlled by $%
PZT_1$ and a feedback photo-electronic circuit, we obtained modes $a_5$ and $%
a_6$. Then, combining modes $a_1$ and $a_5$ on $BS_2$ and $a_4$ and
$a_6$ on $BS_3$ the final four output modes $b_i$ ($i=1,2,3,4$) were
produced. It has been theoretically demonstrated in Ref.[8] (See
Eqs.(4) of Ref.[8]) that the four modes $b_i$ are in the cluster
state if the interfering phase
difference of $a_1$ and $a_5$ is $0$ and that of $a_4$ and $a_6$ is $\pi /2$%
, however they are in the quadripartite GHZ state if both phase
differences are controlled at $0$. The calculated correlation
variances of quadrature components of modes $b_i$ for cluster and
GHZ state are respectively

\begin{eqnarray}
\left\langle \Delta ^2(Y_{b1}^C-Y_{b2}^C)\right\rangle
&=&\left\langle
\Delta ^2(X_{b3}^C-X_{b4}^C)\right\rangle =\frac 12e^{-2r}, \\
\left\langle \Delta ^2(X_{b1}^C+X_{b2}^C+g_3^CX_{b3}^C)\right\rangle  &=&%
\frac{(g_3^C)^2-4g_3^C+4}{16}e^{2r}+\frac{3(g_3^C)^2+4g_3^C+4}{16}e^{-2r},
\nonumber \\
\left\langle \Delta ^2(-g_2^CY_{b2}^C+Y_{b3}^C+Y_{b4}^C)\right\rangle  &=&%
\frac{(g_2^C)^2-4g_2^C+4}{16}e^{2r}+\frac{3(g_2^C)^2+4g_2^C+4}{16}e^{-2r},
\nonumber \\
\left\langle \Delta ^2(g_1^CX_{b1}^C+X_{b2}^C+2X_{b3}^C)\right\rangle  &=&%
\frac{3(g_1^C)^2-6g_1^C+3}{16}e^{2r}+\frac{(g_1^C)^2+6g_1^C+17}{16}e^{-2r},
\nonumber \\
\left\langle \Delta ^2(-2Y_{b2}^C+Y_{b3}^C+g_4^CY_{b4}^C)\right\rangle  &=&%
\frac{3(g_4^C)^2-6g_4^C+3}{16}e^{2r}+\frac{(g_4^C)^2+6g_4^C+17}{16}e^{-2r},
\nonumber
\end{eqnarray}
and

\begin{eqnarray}
\left\langle \Delta
^2(X_{b1}^G+X_{b2}^G+g_3^GX_{b3}^G+g_4^GX_{b4}^G)\right\rangle &=&\frac{%
(2-g_3^G-g_4^G)^2+2(g_3^G-g_4^G)^2}{16}e^{2r}+\frac{(2+g_3^G+g_4^G)^2}{16}%
e^{-2r}, \\
\left\langle \Delta
^2(g_1^GX_{b1}^G+X_{b2}^G+X_{b3}^G+g_4^GX_{b4}^G)\right\rangle &=&\frac{%
(2-g_1^G-g_4^G)^2+2(g_1^G-g_4^G)^2}{16}e^{2r}+\frac{(2+g_1^G+g_4^G)^2}{16}%
e^{-2r}  \nonumber \\
\left\langle \Delta
^2(g_1^GX_{b1}^G+g_2^GX_{b2}^G+X_{b3}^G+X_{b4}^G)\right\rangle &=&\frac{%
(2-g_1^G-g_2^G)^2+2(g_1^G-g_2^G)^2}{16}e^{2r}+\frac{(2+g_1^G+g_2^G)^2}{16}%
e^{-2r},  \nonumber \\
\left\langle \Delta ^2(Y_{b1}^G-Y_{b2}^G)\right\rangle
&=&\left\langle \Delta ^2(Y_{b2}^G-Y_{b3}^G)\right\rangle
=\left\langle \Delta ^2(Y_{b3}^G-Y_{b4}^G)\right\rangle =\frac
12e^{-2r}.  \nonumber
\end{eqnarray}

The upperscript $C$ and $G$ designate the cluster and GHZ state
respectively. $g_i^C$ and $g_i^G$ are the gain factor (arbitrary
real parameter). In experiments we may adjust the electronic gains
of photocurrents to optimize the correlation variances. Calculating
the minimum values of the expressions in Eqs.(2) and Eqs.(3) we got
the optimized gain factors

\begin{eqnarray}
g_{opt1}^C &=&g_{opt4}^C=\frac{3e^{4r}-3}{3e^{4r}+1}, \\
g_{opt2}^C &=&g_{opt3}^C=\frac{2e^{4r}-2}{e^{4r}+3},
\end{eqnarray}
for cluster state and

\begin{equation}
g_{opt}^G=g_{opt1}^G=g_{opt2}^G=g_{opt3}^G=g_{opt4}^G=\frac{e^{4r}-1}{%
e^{4r}+1},
\end{equation}
for GHZ state, respectively.

The four modes $b_i$ and the four local oscillation beams $LO_i$ at
$1080nm$ deriving from the pump laser are sent to the four sets of
the balanced-homodyne-detectors ($BHD_{1-4}$), respectively, for
measuring the fluctuation variances of the amplitude or phase
quadratures of mode $b_i$. The measured photocurrent variances of
each mode are combined by the positive ($+$) or negative ($-$) power
combiner in different way and then are sent to a spectrum analyzer
(SA) for the detection and record of the desired variety correlation
variances.

The experimentally measured squeezing degrees of the output fields
from NOPA1 and NOPA2 equal to $3.50\pm 0.07dB$ below the SNL (the
corresponding squeezing parameter $r$ equals to $0.402\pm 0.012$).
During the measurements the pump power of NOPAs at $540nm$
wavelength is $\sim 200mW$ below the oscillation threshold of
$255mW$ and the intensity of the injected signal at $1080nm$ is
$10mW$.

Adjusting the electronic gains to the optimal values the measured
correlation variances of $\left\langle \Delta
^2(X_{b1}^C+X_{b2}^C+g_{opt2}^CX_{b3}^C)\right\rangle $,
$\left\langle \Delta ^2(X_{b3}^C-X_{b4}^C)\right\rangle $,
$\left\langle \Delta ^2(Y_{b1}^C-Y_{b2}^C)\right\rangle $,
$\left\langle \Delta
^2(-g_{opt2}^CY_{b2}^C+Y_{b3}^C+Y_{b4}^C)\right\rangle $,
$\left\langle
\Delta ^2(g_{opt1}^CX_{b1}^C+X_{b2}^C+2X_{b3}^C)\right\rangle $ and $%
\left\langle \Delta
^2(-2Y_{b2}^C+Y_{b3}^C+g_{opt1}^CY_{b4}^C)\right\rangle $ for the
cluster state are $1.09\pm 0.08dB$, $1.20\pm 0.08dB$, $1.26\pm
0.05dB $, $0.97\pm 0.06dB$, $1.19\pm 0.08dB$ and $1.15\pm 0.07dB$
below the SNL, respectively. The variances of $\left\langle \Delta
^2(X_{b1}^C+X_{b2}^C+g_{opt2}^CX_{b3}^C)\right\rangle $ and
$\left\langle \Delta ^2(Y_{b1}^C-Y_{b2}^C)\right\rangle $ are shown
in Fig.2(a) and (b), and others are not presented for saving the
length of the paper. The measured correlation variances of
$\left\langle \Delta ^2(Y_{b1}^G-Y_{b2}^G)\right\rangle $,
$\left\langle \Delta
^2(X_{b1}^G+X_{b2}^G+g_{opt}^GX_{b3}^G+g_{opt}^GX_{b4}^G)\right\rangle $, $%
\left\langle \Delta ^2(Y_{b2}^G-Y_{b3}^G)\right\rangle $,
$\left\langle \Delta
^2(g_{opt}^GX_{b1}^G+X_{b2}^G+X_{b3}^G+g_{opt}^GX_{b4}^G)\right\rangle $, $%
\left\langle \Delta ^2(Y_{b3}^G-Y_{b4}^G)\right\rangle $, and
$\left\langle \Delta
^2(g_{opt}^GX_{b1}^G+g_{opt}^GX_{b2}^G+X_{b3}^G+X_{b4}^G)\right\rangle
$ for
the GHZ state are $1.20\pm 0.04dB$, $1.18\pm 0.07dB$, $1.16\pm 0.07dB$, $%
1.08\pm 0.08dB$, $1.29\pm 0.09dB$ and $1.07\pm 0.06dB$ below the
SNL, respectively. The variances of $\left\langle \Delta
^2(Y_{b1}^G-Y_{b2}^G)\right\rangle $ and $\left\langle \Delta
^2(X_{b1}^G+X_{b2}^G+g_{opt}^GX_{b3}^G+g_{opt}^GX_{b4}^G)\right\rangle
$ are shown in Fig.3 (a) and (b), and the remainders are omitted
also.

Based on the same method of concluding the full inseparability
criteria of multipartite CV GHZ entanglement in Ref.[12] we derived
the sufficient requirements of full inseparability for the
quadripartite cluster state. When the following three inequalities
are satisfied simultaneously the four submodes $b_i$ are in a fully
inseparable cluster entangled state:

\begin{eqnarray}
I^C\text{ \quad }\left\langle \Delta
^2(Y_{b1}^C-Y_{b2}^C)\right\rangle +\left\langle \Delta
^2(X_{b1}^C+X_{b2}^C+g_{opt2}^CX_{b3}^C)\right\rangle
\\
II^C\text{ \quad }\left\langle \Delta
^2(X_{b3}^C-X_{b4}^C)\right\rangle +\left\langle \Delta
^2(-g_{opt2}^CY_{b2}^C+Y_{b3}^C+Y_{b4}^C)\right\rangle
\nonumber \\
III^C\text{ \quad }\left\langle \Delta
^2(g_{opt1}^CX_{b1}^C+X_{b2}^C+2X_{b3}^C)\right\rangle +\left\langle
\Delta ^2(-2Y_{b2}^C+Y_{b3}^C+g_{opt1}^CY_{b4}^C)\right\rangle
\nonumber
\end{eqnarray}
Substituting the measured correlation variances into the left sides of $I^C$%
, $II^C$ and $III^C$ in Eq.(7) we have $I^C=0.828\pm 0.014$,
$II^C=0.845\pm 0.018$ and $III^C=1.936\pm 0.020$, all of them are
smaller than the normalized SNL. It means the obtained modes are in
a fully inseparable cluster entangled state.

From Ref.[12] we can directly write out the criteria of the full
inseparability for quadripartite GHZ state:

\begin{eqnarray}
I^G\text{ \quad }\left\langle \Delta
^2(Y_{b1}^G-Y_{b2}^G)\right\rangle +\left\langle \Delta
^2(X_{b1}^G+X_{b2}^G+g_{opt}^GX_{b3}^G+g_{opt}^GX_{b4}^G)\right\rangle
\\
II^G\text{ \quad }\left\langle \Delta
^2(Y_{b2}^G-Y_{b3}^G)\right\rangle +\left\langle \Delta
^2(g_{opt}^GX_{b1}^G+X_{b2}^G+X_{b3}^G+g_{opt}^GX_{b4}^G)\right\rangle
\nonumber \\
III^G\text{ \quad }\left\langle \Delta
^2(Y_{b3}^G-Y_{b4}^G)\right\rangle +\left\langle \Delta
^2(g_{opt}^GX_{b1}^G+g_{opt}^GX_{b2}^G+X_{b3}^G+X_{b4}^G)\right\rangle
\nonumber
\end{eqnarray}
Similarly we get $I^G=0.836\pm 0.016$, $II^G=0.849\pm 0.014$ and $%
III^G=0.840\pm 0.020$ from the experimental results. It shows that
all three inequalities in Eq.(8) are satisfied and thus the
quadripartite GHZ entanglement of the four optical modes are
demonstrated experimentally.

For achieving the measurements of a variety of the correlation
variances under same experimental conditions the pump laser has to
operate stably in the whole experimental process. The resonating
frequencies of the laser and the two NOPAs must be locked in a
longer term. Besides, $9$ sets of the phase-locking system are
respectively used for locking the relative phases between the pump
laser and the injected signal beam of the two NOPAs to $\pi $, the
phase difference between $b_i$ and $LO_i$ in the four sets of
$BHD_i$ to $0$ (for amplitude measurement) or $\pi /2$ (for phase
measurement), and
the interference phases between two combined beams on $BS_1$, $BS_2$ and $%
BS_3$ to $0$ or $\pi /2$ according to the different requirements for
generating cluster or GHZ state mentioned above. The standard
side-band frequency-locking\cite{eighteen} and the interference
feedback\cite{nineteen} technologies are utilized in the frequency
and the relative phase locking, respectively.

For conclusion, we have experimentally produced the CV quadripartite
cluster and GHZ entangled states for the first time to the best of
our knowledge. The investigation has demonstrated the possibility to
generate and manipulate two types of CV quadripartite entanglement
using two-mode squeezed states and linear optics, which is the
necessary base for realizing quantum computation and more
complicated quantum communication network.

Acknowledgments: This work was supported by the National Natural
Science Foundation of China(No.60238010).

*Email: jiaxj@sxu.edu.cn

Captions of figures:

Fig.1 Schematic of the experimental setup. Nd:YAP/KTP-laser source; NOPA$%
_{1-2}$-nondegenerate optical parametric amplification;
P$_{1-2}$-$\lambda
/2 $ wave plate; PBS$_{1-2}$-polarizing optical beamsplitter; PZT$_{1-3}$%
-piezo-electric transducer; BS$_{1-3}$-50\% optical beamsplitter; LO$_{1-4}$%
-local oscillation beam; BHD$_{1-4}$-balanced-homodyne-detectors; $+/-$%
-positive/negative power combiner; SA-spectrum analyzer

Fig.2 The measured the measured correlation variances of the cluster
state at 2MHz as a function of time. (a): $\left\langle \Delta
^2(X_{b1}^C+X_{b2}^C+g_{opt2}^CX_{b3}^C)\right\rangle $, (b):
$\left\langle \Delta ^2(Y_{b1}^C-Y_{b2}^C)\right\rangle $. 1, the
shot noise limit (SNL); 2, The correlation noise power. The
measurement parameters of SA: RBW(Resolution Band Width)-30kHz;
VBW(Video Band Width)-30Hz.

Fig.3 The measured the measured correlation variances of the GHZ
state at 2MHz as a function of time. (a): $\left\langle \Delta
^2(Y_{b1}^G-Y_{b2}^G)\right\rangle $, (b): $\left\langle \Delta
^2(X_{b1}^G+X_{b2}^G+g_{opt}^GX_{b3}^G+g_{opt}^GX_{b4}^G)\right\rangle
$. 1, SNL; 2, The correlation noise power. The measurement
parameters of SA are same as Fig.2.

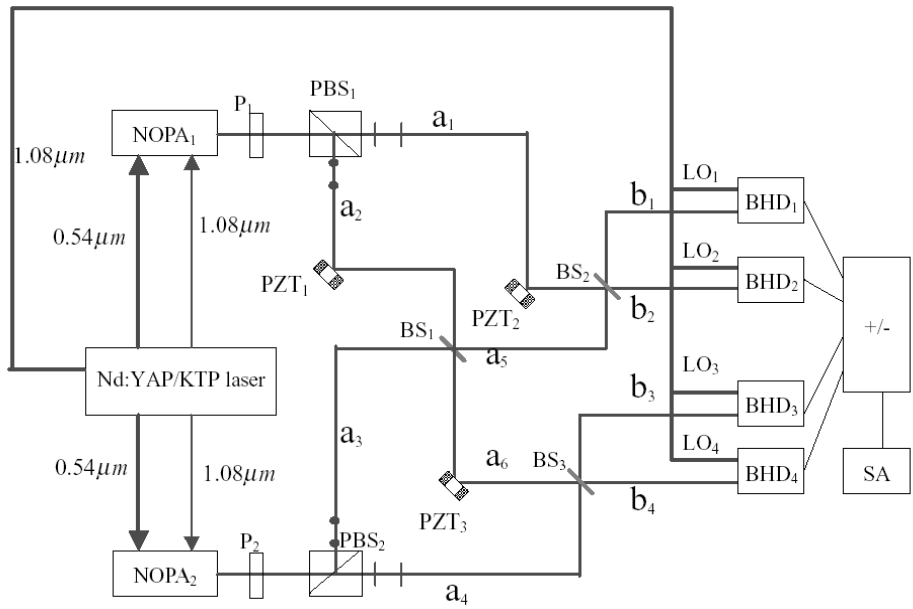
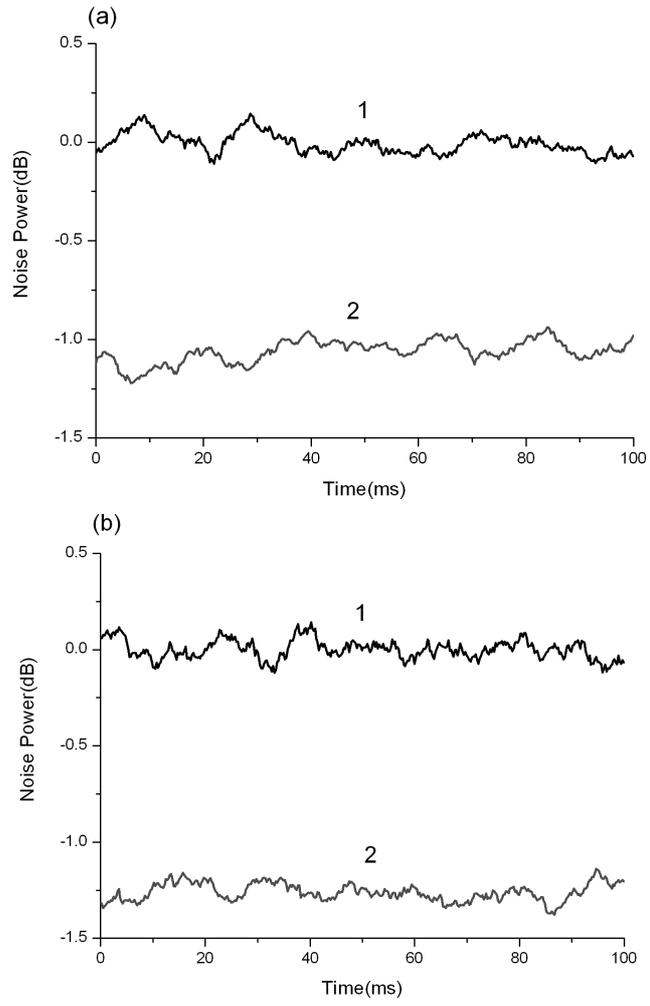
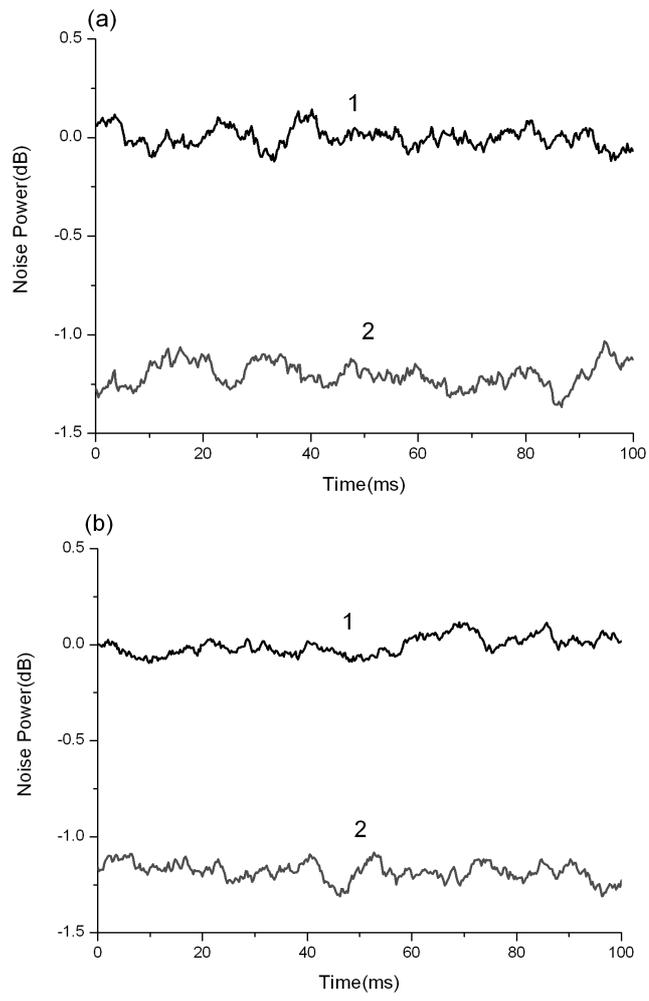
\end{document}